\documentclass[twocolumn]{revtex4}

\usepackage{amsmath}
\usepackage{amssymb}

\makeatletter
\usepackage{epsfig}

\makeatother

\begin{document}

\title{Memory Effect in Upper Bound of Heat Flux Induced by Quantum Fluctuations}

\author{T. Koide}

\address{Instituto de F\'{i}sica, Universidade Federal do Rio de Janeiro, C.P.
68528, 21941-972, Rio de Janeiro, Brazil }
\begin{abstract}

Thermodynamic behaviors in a quantum Brownian motion coupled to 
a classical heat bath is studied. 
We then define a heat operator by generalizing the stochastic energetics 
and show the energy balance (first law) and the upper bound of the expectation value 
of the heat operator (second law). 
We further find that this upper bound depends on the memory effect 
induced by quantum fluctuations and hence the maximum extractable work 
can be qualitatively modified in quantum thermodynamics.

\end{abstract}

\maketitle


\section{Introduction}

The accelerating development in nanotechnologies enables us to 
access individual thermal random processes at microscopic scales. 
External operations to these systems cause various responses which are 
understood through quantities such as energy, work and heat. 
However we cannot directly apply thermodynamics to these quantities 
because the typical scale of the systems is very small and the effect of thermal fluctuations is not negligible. 
There is no established theory to describe general fluctuating systems thermodynamically \cite{small}.
On the other hand, such a system is often modeled as a Brownian motion \cite{hanngi-review} 
and then the behaviors 
can be interpreted thermodynamically by using the stochastic energetics (SE) \cite{se-book}.

In this theory, energy, work and heat are represented by the variables of the Brownian particles, and  
we can show that 
the energy balance is satisfied and the expectation value of the heat flux has an upper bound.
The former corresponds to the first law and the latter the second law in thermodynamics, respectively. 
The various applications of SE are discussed in Ref.\ \cite{se-book}. 
The prediction of SE is experimentally confirmed by analyzing extracted works from a microscopic heat engine 
\cite{blickle}. 
Although this theory is generalized to relativistic systems \cite{kk-rel} and the Poisson noise \cite{kanazawa}, 
the applications are still limited to classical systems \cite{ghosh}.

On the other hand, the emergence of thermodynamic behaviors 
in quantum systems is another intriguing problem \cite{qt-review,brasil}.
In particular, it is interesting to ask whether thermodynamic behaviors are qualitatively modified 
by quantum fluctuations \cite{uzdin}.
For example, the maximum extractable work may be
limited by quantum coherence in a small system \cite{horo2013}.
To identify modified behaviors by quantum fluctuations, it is important to formulate 
a theory which has a well-defined classical limit \cite{qt-cla}.

In this work, we study a formulation of quantum thermodynamics by 
generalizing SE to a quantum Brownian motion coupled to a classical heat bath \cite{qbm}. 
Our model is characterized by stochastic differential equations of the position and momentum operators 
of the quantum Brownian particle. 
Then, the behaviors of other operators are determined from the two equations 
by employing a differential with respect to operators in the quantum analysis \cite{qa}.
We then define a heat operator, showing properties corresponding to the first and second laws in thermodynamics.
Our theory has a well-defined classical limit and reproduces the results of the classical SE.
Moreover we find that the behavior of the heat is qualitatively modified from the classical 
one by quantum fluctuations, affecting the maximum extractable work in quantum heat engines.

This paper is organized as follows. In Sec.\ \ref{sec:model}, a model of a quantum open system 
based on the quantum Brownian motion is developed.
In Sec.\ \ref{sec:qse}, we define thermodynamic properties of this model 
by extending SE and show the modification of the second law by the effect of quantum fluctuations. 
Section \ref{sec:final} is devoted to concluding remarks and discussions.

\section{Definition of Model} \label{sec:model}

Our model of a quantum open system is characterized by stochastic differential equations (SDE's) for a position operator $\hat{x}_t$ and 
a momentum operator $\hat{p}_t$ of a quantum Brownian particle, which are defined by 
\begin{subequations} 
\begin{eqnarray}
 &  & \hspace*{-1cm}d\hat{x}_{t}=\frac{1}{m}\hat{p}_{t}dt, \label{qsde1}\\
 &  & \hspace*{-1cm}d\hat{p}_{t}=-\frac{\nu}{m}\hat{p}_{t}dt
- V^{(1)} (\hat{x}_{t},\lambda_t)dt + \sqrt{2\nu k_{B}T}dB_{t}, \label{qsde2}
\end{eqnarray} 
\label{qsdes} 
\end{subequations} 
\hspace*{-0.1cm}where $k_B$, $m$, $T$ and $\nu$ are the Boltzmann constant, mass, temperature of a heat bath and dissipative coefficient, respectively. 
The external potential $V$ depends on an external parameter $\lambda_t$ 
and $V^{(n)} (x,\lambda_t ) \equiv \partial^n_x V(x,\lambda_t)$. 
The symbol $\hat{~~}$ denotes operator.

These equations can be obtained 
from a microscopic dynamics by using, for example, the projection operator technique and the Markov limit \cite{qbm,breuer-book,zwanzig}.
Note that, because we consider a dissipative system, there is no Lagrangian which reproduces this system, and thus 
$\hat{x}_t$ and $\hat{p}_t$ are not canonical variables in general.
However, to maintain the notation in the classical Brownian motion, we still call $\hat{p}_t$, which is defined by Eq.\ (\ref{qsde1}), 
momentum operator.

The last term $\sqrt{2\nu k_{B}T}dB_{t}$, which is called noise term, represents thermal fluctuations 
induced by the interaction with a heat bath and shows a stochastic behavior. 
In principle, this term also can be replaced by an operator, but 
the definition of operators in the stochastic calculus is not well-understood. 
Thus we here treat the noise term as a stochastic c-number, that is, the increment of the standard Wiener process 
defined by the following correlation properties \cite{gardiner}, 
\begin{eqnarray}
E[dB_{t}] =  0,\ \ \  E[(dB_{t})^{2}] =  dt,
\end{eqnarray}
Other second order correlations vanish.
We assume the existence of an appropriate probability space ($\sigma$-algebra) 
for $\hat{x}_t$ and $\hat{p}_t$ \cite{gardiner,breuer-book}. 
As we will see later, because of this idealization, the heat bath behaves as a classical degree of freedom.

In this formulation, the behaviors of other operators should be obtained from the above two SDE's.
To implement this systematically, we define a differential in terms of operators applying 
the quantum analysis (QA) \cite{qa}.

\subsection{Quantum analysis}

QA was proposed to expand the functions of operators systematically 
and has been applied to various problems in quantum mechanics and quantum statistical mechanics.
For example, the expansion of
the S-matrix, the Baker-Campbell-Hausdorff formula and the linear response theory 
can be regarded as the operator Taylor expansion in QA \cite{qa}.

Let us consider $f(\hat{A})$ where $f(x)$ is a smooth function of $x$.
Then the operator differential with respect to $\hat{A}$ is expressed 
by $(df/d\hat{A})$, and introduced through the following equation,  
\begin{equation}
f(\hat{A}+h\hat{C})-f(\hat{A})=\left(\frac{df}{d\hat{A}}\right)h\hat{C}+O(h^2),\label{dff1}
\end{equation}
where $h$ is a small c-number and $\hat{C}$ is another operator which is in general not commutable with $\hat{A}$, 
$[\hat{A},\hat{C}] \neq 0$.
Note that the value of the differential depends on the operator 
$\hat{C}$ and thus $(df/d\hat{A})$ is a hyper operator.

In QA, this operator differential is defined by  
\begin{equation}
\left( \frac{df}{d\hat{A}} \right) =\int_{0}^{1}d\lambda f^{(1)}(\hat{A}-\lambda\delta_{A}),\label{qaf1}
\end{equation}
where  
$\delta_{A}=[\hat{A},~~]$.

The advantage of this definition is that the operator Taylor expansion is expressed in the following 
simple form, 
\begin{eqnarray}
f(\hat{A}+\hat{C})=f(\hat{A})+\sum_{n=1}^{\infty}\frac{1}{n!}\left( \frac{d^{n}f}{d\hat{A}^{n}} \right) 
\hat{C}^{n},
\end{eqnarray}
where 
\begin{equation}
\hspace*{-0.5cm} \left( \frac{d^{n}f}{d\hat{A}^{n}} \right) =n!\int_{0}^{1}d\lambda_{1}\cdots\int_{0}^{\lambda_{n-1}}d\lambda_{n}f^{(n)}(\hat{A}-\sum_{i=1}^{n}\lambda_{i}\delta_{A}^{(i)}),
\end{equation}
with 
\begin{equation}
\delta_{A}^{(i)}\hat{C}^{n}=\hat{C}^{n-i}(\delta_{A}\hat{C})\hat{C}^{i-1}.
\end{equation}

Moreover, when $\hat{A}_t$ is a function of a c-number $t$,
we have 
\begin{eqnarray}
\frac{df(\hat{A}_{t})}{dt} = \left( \frac{df}{d\hat{A}_{t}} \right) \frac{d\hat{A}_{t}}{dt}.
\end{eqnarray}

Several useful relations for $\delta_{A}$ are summarized as 
\begin{subequations}
\begin{eqnarray}
&& {[\hat{A},\delta_{A}]} =  0,\ \ \ \ \ \ f(\hat{A}-\delta_{A})\hat{C}  = \hat{C}f(\hat{A}), \\
&& \delta_{A}\hat{C} =  -\delta_{C}\hat{A},\ \ \ e^{a\delta_{A}}\hat{C} = e^{a\hat{A}}\hat{C}e^{-a\hat{A}}.
\end{eqnarray}
\end{subequations}

Let us apply the above definitions to an operator given by the following SDE, 
\begin{eqnarray}
d\hat{A}_{t}=\hat{L}_{t}dt+\sqrt{2\nu T}d{B}_{t}, \label{sde-op}
\end{eqnarray}
where $d\hat{A}_{t}=\hat{A}_{t+dt}-\hat{A}_{t}$.
Using the operator Taylor expansion for $f(\hat{A}_{t} + d\hat{A}_t)$ and Eq. (\ref{sde-op}), 
we find 
\begin{eqnarray}
df(\hat{A}_{t}) 
&=& 
\left[ \int^1_0 d\lambda f^{(1)} (\hat{A}_t - \lambda \delta_{A_t}) \hat{L}_t + \nu T f^{(2)}(\hat{A}_t )\right] dt \nonumber \\
&& + \sqrt{2\nu T} f^{(1)} (\hat{A}_t)\circ_i dB_t \\
&=& \left( \frac{df(\hat{A}_{t})}{d\hat{A}_{t}} \right) \circ_{s}d\hat{A}_{t}.\label{i-s-tra}
\end{eqnarray}
Here the terms of $O(dt^{3/2})$ are dropped.
The products $\circ_i$ and $\circ_s$ are, respectively, given by 
the Ito definition, 
\begin{eqnarray}
f(\hat{A}_{t}) \circ_{i}dB_{t}\equiv f(\hat{A}_{t})(B_{t+dt}-B_{t}),
\end{eqnarray} 
and the Stratonovich definition,
\begin{eqnarray}
f(\hat{A}_{t}) \circ_{s}dB_{t}\equiv f(\hat{A}_{t+dt/2})(B_{t+dt}-B_{t}).
\end{eqnarray} 
This result is the operator extension of Ito's lemma in the usual stochastic calculus \cite{gardiner}.

There is a convenient formula satisfied for operators $\hat{A}$
and $d\hat{A}$, which have a constant commutator, $[\hat{A},d\hat{A}]=const$,
\begin{eqnarray}
\left( \frac{df}{d\hat{A}} \right) \circ_{s}d\hat{A}=\left(d\hat{A}-\frac{1}{2}\delta_{dA}\right)\circ_{s}f^{(1)},\label{myformula}
\end{eqnarray}
where $\delta_{A}\circ_{s}\hat{C}=\hat{A}\circ_{s}\hat{C}-\hat{C}\circ_{s}\hat{A}$.

\subsection{Commutation relation}


By applying QA, the differential of the commutator of $\hat{x}_t$ and $\hat{p}_t$ in our model is
\begin{eqnarray}
d[\hat{x}_{t},\hat{p}_{t}]
=-\frac{\nu}{m}dt[\hat{x}_{t},\hat{p}_{t}] + O(dt^{3/2}).
\end{eqnarray}
We consider that  
the quantum Brownian particle starts to interact with the classical heat bath  
at the initial time $t=0$ and thus $[\hat{x_{0}},\hat{p}_{0}]=i\hbar$.
Using this condition, the solution of the above equation is  
\begin{eqnarray}
[\hat{x}_{t},\hat{p}_{t}] = i\hbar e^{-\nu t/m}  \equiv  i\hbar\gamma(t). \label{commu-diss}
\end{eqnarray}

One can see that the commutator vanishes in the asymptotic limit in
time and then $\hat{x}_{t}$ and $\hat{p}_{t}$
behave as classical variables.
This time dependence is the nature of Eq.\ (\ref{qsdes}), and irrelevant to the properties of QA. In fact, 
for the case of $V=0$, we can directly solve Eq.\ (\ref{qsdes}) and confirm that Eq.\ (\ref{commu-diss}) 
is satisfied.

It should be noted that our model is different from Kanai's model where a damping harmonic oscillator is quantized, 
although a similar time-dependent commutator is obtained. In fact, a coupling to a classical heat bath is not considered 
in Kanai's approach \cite{kanai}.

\subsection{Wigner function and equilibrium distribution} \label{sec:wigner-fp}

The above behavior of the commutator indicates that 
our model relaxes toward a classical equilibrium state. 
To see this relaxation, we introduce the Wigner function, 
\begin{equation}
\rho_{W}(x,p,t)=\langle\langle\delta(x-\hat{x}_{t}+\delta_{x_{t}}/2)\delta(p-\hat{p}_{t})\rangle\rangle,
\label{wigner-qa}
\end{equation}
where $\langle\langle\ \ \rangle\rangle$ denotes a double expectations: 
one is for the Wiener process $E[~~]$ and the other for an initial wave function $|\psi_{0}\rangle$,
\begin{eqnarray}
\langle\langle\hat{A}\rangle\rangle
= \langle\psi_{0}|E[\hat{A}]|\psi_{0}\rangle 
= E [ \langle\psi_{0}|\hat{A}|\psi_{0}\rangle] .
\end{eqnarray}
Note that the initial wave function is independent of the Wiener process and 
the order of the quantum and stochastic averages can be exchanged.
The delta function here is defined by the integral form, 
$\delta(x)=\frac{1}{2\pi}\int dke^{ikx}$.

The definition by Eq.\ (\ref{wigner-qa}) is different from the traditional expression
of the Wigner function \cite{wigner}, but still gives the same result. 
One can see from this expression  
that the Wigner function is reduced to the classical phase space distribution in the classical limit.

Using QA, the time derivative of $\rho_{W}(x,p,t)$ is calculated as
\begin{eqnarray}
&& \hspace*{-0.5cm} \lefteqn{\partial_t \rho_{W}(x,p,t)}\nonumber \\
&& \hspace*{-0.5cm} =\left[-\frac{p}{m}\partial_{x}+V^{(1)}(x,\lambda_t)\partial_{p}+\frac{\nu}{m}\partial_{p}p
+\frac{\nu}{\beta} \partial_{p}^{2}\right]\rho_{W}(x,p,t)\nonumber \\
&& \hspace*{-0.5cm}+ \Sigma(x,p,t),\label{qse-fp}
\end{eqnarray}
where $\beta^{-1} = k_B T$ and
\begin{eqnarray}
\hspace*{-0.5cm}
\Sigma(x,p,t) = \sum_{l=1}^{\infty} \frac{V^{(2l+1)}(x,\lambda_t)}{(2l+1)!} \left( -\frac{\hbar^2}{4} \gamma^2(t) \right)^{l} 
\partial^{2l+1}_p
\rho_{W}. \label{Sigma}
\end{eqnarray}

In the vanishing limit of dissipation, $\nu \rightarrow 0$, Eq.\ (\ref{qse-fp}) 
is reduced to the well-known result in quantum mechanics \cite{wigner}.
In the classical limit, $\hbar \rightarrow 0$ and/or in the asymptotic limit in time $t\rightarrow \infty$, 
$\Sigma$ disappears and Eq.\ (\ref{qse-fp}) coincide with the Kramers (Fokker-Planck) equation of 
the classical Brownian motion \cite{se-book}.

The Wigner functions for various quantum open systems are discussed in Ref.\ \cite{open-lindblad1} and 
one of them is the case of a quantum Brownian motion with a noise operator.
Then the Wigner function of this model is the same as Eq.\ (\ref{qse-fp}), replacing the factor $\gamma(t)$ by one. 
However, the definition of the noise operator used there is incomplete to formulate stochastic calculus.

For later discussion, we introduce the solution of the Kramers equation 
by $\rho_{KR}(x,p,t)$. Then $\rho_W(x,p,t) = \rho_{KR}(x,p,t)$ in the classical limit.

The stationary solution of Eq.\ (\ref{qse-fp}) is given by 
\begin{eqnarray}
\lim_{t\rightarrow \infty} \rho_W (x,p,t)=\rho_{eq}(x,p)=\frac{1}{Z_c} e^{- \beta H(x,p,\lambda_{eq})}, \label{rho-lim}
\end{eqnarray}
where $Z_c$ is the partition function, $Z_c = \int d\Gamma e^{-\beta H}$ with the phase volume 
$d\Gamma = dx dp$, and 
\begin{eqnarray}
H(x,p,\lambda_{eq}) = \frac{p^2}{2m} + V(x,\lambda_{eq}),
\end{eqnarray}
with a constant $\lambda_{eq} = \lambda_{t=\infty} $.
This is nothing but the classical equilibrium distribution as is expected from the behavior of 
the commutator.

The Wigner function is not positive definite and thus cannot be interpreted 
as a probability density. 
Instead, it should be interpreted as an integration measure. 
As a matter of fact, we can re-express any expectation values of operators by integrals 
with this measure.
For example, the energy expectation value is rewritten as
\begin{eqnarray}
\langle\langle H(\hat{x}_t, \hat{p}_t,\lambda_{t}) \rangle\rangle =\int d\Gamma \rho_{W}(x,p,t) H(x,p,\lambda_{t}).
\end{eqnarray}

\section{Quantum Stochastic Energetics coupled to classical heat bath} \label{sec:qse}

In the classical SE, the heat absorbed by a Brownian particle 
is defined as the work exerted by the heat bath on the Brownian particle.  
In fact, the interaction between the particle and the bath is
represented by the dissipative term ($-\nu \hat{p}_{t}/m$ in Eq.\ (\ref{qsde2}) in the present model) 
and the noise term ($\sqrt{2\nu T}dB_{t}/dt$). 
The heat {\it absorbed} from the heat bath is equivalent to 
the work {\it exerted} by the heat bath on the Brownian particle, which is, thus, 
defined by the product of a force and an induced displacement \cite{se-book}.

Extending this idea to quantum systems, note that the force and the displacement 
are operators and not commutable in general. 
Here we propose a heat operator as 
\begin{eqnarray}
d\hat{Q}_t  \equiv  \left( d\hat{x}_{t}-\frac{1}{2}\delta_{dx_{t}} \right)\circ_{s}
\left( -\frac{\nu}{m}\hat{p}_{t}+\sqrt{2\nu T}\frac{dB_{t}}{dt} \right).
\end{eqnarray}
The operator $\delta_{dx_t}$ symmetrizes the order 
of the force and the displacement operators.

By using the properties in QA, in particular Eq.\ (\ref{myformula}), 
we can show that the heat operator satisfies the following energy balance, 
\begin{eqnarray}
dH(\hat{x}_t, \hat{p}_t,\lambda_t)=d\hat{Q}_t + d\hat{W}_t. \label{1stlaw}
\end{eqnarray}
Here the work operator exerted by an external force is defined by 
\begin{eqnarray}
d\hat{W}_t \equiv \partial_{\lambda}V(\hat{x}_{t},\lambda_t)\circ_{s}d\lambda_t,
\end{eqnarray}
because the external force changes the form of $V$ through its $\lambda_t$ dependence.
This energy balance (\ref{1stlaw}) corresponds to the 
first law of thermodynamics and 
is equivalent to that in the classical SE except for the difference 
of operators and c-numbers.
Note that the energy balance is satisfied not for ensembles but for operators.

The expectation value of the heat operator has an upper bound. 
To see this, we introduce a function, 
\begin{eqnarray}
\hspace*{-0.5cm} S (t) = S_{SH} (t) + S_{ME}(t),  \label{s_sigma}
\end{eqnarray}
where 
\begin{eqnarray}
&& \hspace*{-1cm} S_{SH} (t) = -k_{B}\int d\Gamma\rho_{W}(x,p,t)\ln|\rho_{W}(x,p,t)|, \\
&& \hspace*{-1cm} S_{ME} (t) = k_B\int^t ds \int d\Gamma \biggl[  \Sigma(x,p,s) \ln |\rho_{W}(x,p,s)|  \nonumber \\
&& \hspace*{-1cm} \left. - \beta \nu \delta^{(\hbar)} \rho_W (x,p,s) 
\left\{\frac{p}{m}+ \beta^{-1}\partial_{p} \ln |\rho_{W}(x,p,s)| \right\}^{2} \right].
\end{eqnarray}
Here $\delta^{(\hbar)} \rho_W (x,p,t) \equiv \rho_W (x,p,t)- \rho_{KR}(x,p,t)$ and represents the modification 
of the phase space distribution by quantum fluctuations. 
The first term $S_{SH}(t)$ is the Shannon entropy calculated by using the Wigner function instead of a probability distribution.
The second term $S_{ME}(t)$ contains the memory effect and thus the behavior 
of $S(t)$ depends on the hysteresis of the evolution. 
Note that $S_{ME}(t)$ is induced by quantum fluctuations 
and thus vanishes in the classical limit, leading to $S(t) = S_{SH}(t)$.

Then we can show the following inequality,  
\begin{eqnarray}
\hspace*{-0.0cm} T\frac{dS}{dt} - \langle\langle\frac{d\hat{Q}_t}{dt} \rangle\rangle
&=& \nu \int d\Gamma \  \rho_{KR}\left\{\frac{p}{m}+ \beta^{-1} \partial_{p} \ln |\rho_{W}| \right\}^{2} \nonumber \\
&\ge& 0. \label{2ndlaw}
\end{eqnarray}
The right hand side on the first line is positive definite and vanishes when $\rho_W = \rho_{eq}$. 
Therefore the upper bound of the expectation value of the heat flux is characterized by the time derivative of $S(t)$. 
This inequality corresponds to the second law of thermodynamics.
As a matter of fact, $S(t)$ can be interpreted as the thermodynamic entropy in equilibrium, 
because 
\begin{eqnarray}
\left. S\right|_{\rho_W = \rho_{eq}} = \left. S_{SH} \right|_{\rho_W = \rho_{eq}} = \frac{\langle \langle \hat{H} \rangle \rangle }{T}
+ k_b \ln Z_c,
\end{eqnarray}
where $Z_c$ is the partition function defined above.

In the classical limit, 
our Wigner function coincides with the phase space distribution $\rho_{KR}$ as is discussed above 
and Eq.\ (\ref{2ndlaw}) is reduced to 
$T dS_{SH}/dt \ge E\left[ dQ_t/dt \right]$ which 
is the result in the classical SE \cite{se-book}. That is, our quantum SE 
has a consistent classical limit for the first and second laws. See also Table \ref{table1} 
for the classical definition of $dQ_t$.

The most important nature of the above result is the appearance of the memory effect 
in $S_{ME}(t)$ induced by quantum fluctuations. 
As a consequence, it is expected that the thermal efficiency of quantum heat engines 
will be different from that of the classical one.
To see this effect formally, let us consider two processes interacting with 
different heat bathes of temperatures $T_l$ and $T_h$ ($T_l < T_h$). 
Applying Eqs.\ (\ref{1stlaw}) and (\ref{2ndlaw}), the work per unit time {\it extracted} by interacting with the heat bath of $T_i$ 
has an upper bound given by
\begin{eqnarray} 
-\frac{ d \langle\langle H \rangle\rangle_i }{dt} + T_i \frac{dS^{i}}{dt}, \label{freeener}
\end{eqnarray} 
where the index $i(=l,h)$ represents a quantity observed in each system of $T_i$.
Combining these and appropriate adiabatic processes, we can construct a cycle and then  
the total work extracted from this cycle $W_{EXT}$ has a following limitation,
\begin{eqnarray}
W_{EXT} \le  T_l \Delta S^l + T_h \Delta S^h, \label{extractable}
\end{eqnarray}
where $\Delta S^i$ is the time integration of $dS^i (t)/dt$ for a period of the interaction 
with the heat bath of $T_i$.
The right hand side depends on the memory effect. If this gives a negative contribution, 
the efficiency can be smaller than that of thermodynamics.

\begin{center}
\begin{table}

\begin{tabular}[t]{|c|ccc|}
\hline
& heat & & upper bound  \\
\hline 
classical  & $dx_t \circ_s F (p_t, dB_t) $ & & $TdS_{SH}/dt $ \\
quantum & 
$\left( d\hat{x}_t - \frac{1}{2} \delta_{dx_t}\right) \circ_s F (\hat{p}_t, dB_t) $  & & $Td(S_{SH} + S_{ME})/dt$ \\
\hline
\end{tabular}

\caption{Comparison of the classical and quantum SE. 
We introduce $F (p_t, dB_t)\equiv -\frac{\nu}{m}p_t + \sqrt{2\nu k_B T} \frac{dB_t}{dt}$.}
\label{table1}

\end{table}
\end{center}

\section{Concluding remarks and discussions} \label{sec:final}

In this work, we considered thermodynamic behaviors in a quantum Brownian motion coupled to 
a classical heat bath. 
We then defined a heat operator by generalizing the stochastic energetics 
and showed the energy balance (first law) and the upper bound of the expectation value 
of the heat operator (second law).
Our theory has a well-defined classical limit and reproduces the results of the classical SE.

We observe additional restrictions for observables when the classical SE is generalized to quantum systems.
In fact, the commutation relations of the heat operator are calculated as   
\begin{subequations}  
\begin{eqnarray}
&&\hspace*{-0.5cm} [\hat{p}_t, d\hat{Q}_t]_{i} 
\equiv  \hat{p}_t \circ_i d\hat{Q}_t - d\hat{Q}_t \circ_i \hat{p}_t = 0, \\ 
&&\hspace*{-0.5cm} \protect{[\hat{x}_t, d\hat{Q}_t]_i}  
= \frac{2i\hbar}{m} \left\{ \dot{\gamma}(t)\hat{p}_t dt + \gamma(t) \sqrt{\frac{\nu}{2\beta}} dB_t \right\} ,
\end{eqnarray}
\end{subequations}
where $ \dot{\gamma}(t) = \partial_t \gamma (t)$. From the second equation, we can show 
\begin{eqnarray}
(\Delta x_t) \left( \Delta \frac{dQ_t}{dt} \right) \ge \frac{\hbar}{m}  | \dot{\gamma} (t)\langle \langle \hat{p}_t \rangle \rangle| ,
\end{eqnarray}
where $\Delta A = \sqrt{ \langle \langle \hat{A}^2  \rangle \rangle - (\langle \langle \hat{A}  \rangle \rangle)^2}$.
Therefore, there will exist a limitation for the simultaneous measurement of quantum thermodynamic quantities.

To generalize this approach to a system coupled to a quantum heat bath, 
the noise term will be replaced by an operator. 
In fact, an operator equation of a quantum Brownian motion may be derived from an 
underlying microscopic theory by employing systematic coarse-grainings procedures such as  
the projection operator technique, the influence functional method and so on \cite{qbm,zwanzig,breuer-book}.
Then the derived operator equation contains a term identified with noise.
This term is expected to show stochastic behavior by taking the Markov limit, 
but there is no proof so far and the properties of such an operator have not yet been well understood \cite{arimitu,koidepom}. 
Thus the introduction of a noise operator is not a trivial task \cite{qbm,hudson}.
We are, in particular, interested in whether completely positive maps can be realized 
by introducing a noise operator.

Because of the classical treatment of the heat bath, 
this model describes only a part of quantum fluctuations. 
Nevertheless, we still observed that quantum fluctuations can modify thermodynamic behaviors qualitatively. 
In fact, we found the appearance of the memory effect in the upper bound, which 
can modify the qualitative nature of the maximum extractable work in quantum heat engines. 
This result resembles Ref.\ \cite{horo2013} where  
a limitation on maximum extractable work in a quantum small system is discussed  
by analyzing the modification of the Helmholtz free energy in the quantum information theory.  
As is seen from Eq.\ (\ref{freeener}), we can introduce another free energy 
characterizing the work limitation as 
$\tilde{F} = \langle \langle H \rangle \rangle - T S$, which coincides 
with the Helmholtz free energy for quasi-static processes because of the memory effect in $S$.
See also the different conclusion in Ref.\ \cite{uzdin} for the effect of quantum fluctuations in quantum heat engines.

Note that a possible entanglement between a Brownian particle and 
a heat bath is not included in the present model.
To consider this effect, of course, we need to introduce a noise operator which 
has a well-defined stochastic behaviors.
There is however another problem to deal with such an entanglement.
In the microscopic derivations of the classical and quantum Brownian motions, 
it is normally assumed that there is no correlation between the system and bath density matrices, at least initially \cite{zwanzig,breuer-book}. 
Thus there exists a limitation in the discussion of the system-bath entanglement in such a dynamics.

The memory effect contains terms which have higher order derivatives in momentum and thus 
may survive even near equilibrium for relativistic systems which have 
an energy dispersion $\sqrt{p^2 + m^2}$ \cite{kk-rel}. 
Then it will be interesting to consider the application of quantum thermodynamics 
to the physics of graphene.

\vspace{0.5cm}

The author acknowledges E.\ Brigatti, R.\ Kurita and 
the ICE group of the institute of physics for fruitful discussions and comments. 
This work is financially supported by CNPq.

\end{document}